\documentclass[11pt,fleqn]{article}
\usepackage{psfig}
\title{Learning of  Atomic Physics and Quantum Mechanics : \\ Which should Begin First}
\author{Chen Qin$^1$ and Zotin K.-H. Chu$^2$} % \cite{Chu:2010}} %\thanks{Present Address :
\date{$^{1}$Physics Department, Xinjiang Normal University, Urumqi
830054, China and \\
$^2$P.O. Box 39, Distribution Unit, XiHong Road, Urumqi 830000,
China}
\begin{document}           % End of preamble and beginning of text.
\maketitle
\begin{abstract}
What are the differences and similarities between atomic-physics
studies at different peoples (Han, Kazak and Uygur perples in the
same university) across Xinjiang (a far-west district in PR China
which is a border for previous USSR and Kazak)? In this short
report we focus on issues relating to the learning style of
different-people students to pass the atomic physics course in
physics department even the quantum mechanics course has not been
taken before. \newline

\end{abstract}
%\pacs{62.65.+k, 71.23.An, 91.60.Lj, 47.56.+r}
%03.75.Hh, 03.75.Kk, 05.30.Jp}
%03.75.Hh, 03.75.Lm, 05.70.-a}
%%\end{titlepage}      %%\twocolumn     %%\nopagebreak   %
\oddsidemargin=1mm
\doublerulesep=6.8mm    %
\baselineskip=6.8mm
\bibliographystyle{plain}
\section{Introduction}
Atomic physics is the academic subject that studies the inner
workings of the atom as well as the interactions of the
environment and atoms. It remains one of the most important
testing grounds for quantum theory, and is therefore still an area
of active research, both for its contribution to fundamental
physics and to technology. Furthermore, many other branches of
science rely heavily on atomic physics. The following  gives a few
examples: Astrophysics, plasma physics, atmospheric physics, solid
state physics, chemical physics and radiation physics for physical
science. Chemistry (analysis, reaction rates), biology (molecular
structure, physiology), materials science, energy research, fusion
studies are for other sciences.  Lasers, X-ray technology, NMR,
pollution detection, medical applications of devices (lasers, NMR
etc.) are for direct applications.
\newline
Unfortunately the study of physics is not seen as a good option
today in China (PR China). The worst holds for the far-west of PR
China (PRC), like Xinjiang district now$^{1-2}$. The area of
Xinjiang district is around 1/7 of the total area of PRC. The
capital city of the Xinjiang district is Urumqi (where around 2.5
millions of Han, Uygur, Kazak peoples live therein)  and there are
limited educational resources like universities and academic
institutes (the most important ones belong to the China Academy of
Science or CAS) up to now$^{1-2}$. Most of those students studying
in physics prefer choosing material, condensed matter and optical
physics as their major. Nowadays even a few of the best (physics)
postgraduates (holding MSc degrees) can only got teaching jobs in
high schools in Urumqi.
\newline The decrease of interest in the study of atomic physics
(an undergraduate (UG) course in PR China's university) has, as we
have already surveyed, two important causes. The first reason is
that the job market as well as the chances for postgraduate study
in Xinjiang (PR China) actually requires few
atomic-physicists$^1$. The second is that atomic physics, as an
important field of knowledge, seems to have reached a plateau, and
no immediate revolutionary developments in high technology are
foreseen, at least for Xinjiang District in PR China.\newline Our
paper is structured as follows: in Section 2 we give some data on
the verification of atomic-physics teaching in Xinjiang Normal
University; in Section 3 we describe the details of the
comparative study between different-peoples UGs and in the last
section  we discuss the results. The presentations seem to us
significant and interesting, not just in relation to the
undergraduates that have been tested, but for the current system
of atomic-physics teaching, and future reform.
\section{Verification of Atomic-Physics Learning}
We shall present an  evaluation system for the primary learning of
atomic physics (UG courses) in Xinjiang Normal University (XJNU).
This course is offered for year-2 Han-people UGs but for
Other-people UGs, they can only study this course at year-3. The
latter is due to the poor listening as well as the writing of
Mandarin (Putonghua or Han-language) and Chinese, not to mention
the speaking of Mandarin$^3$. After entering the university
(XJNU), all Other-people UGs should learn Mandarin (speaking and
listening) and Chinese (writing) at year-1 academic year$^3$. The
remaining parts of the time during year-1 all Other-people UGs
should also try to catch up those prerequisites, like basic
mathematics.  Note that there exists a task of training ethnic
bilingual teachers (mainly for high schools) in Xinjiang around
2004 for some universities and institutes therein at Urumqi. All
UGs have the well-known (Chinese) textbook : Atomic Physics
written and edited by Ch'u Sheng-Lin$^4$. We like to remind the
readers that Dr. Ch'u S-L got PhD from Univ. Chicago in 1935
(under the thesis supervisor : Prof. Depmster, AJ : working in the
high frequency spark field)$^5$.\newline Most of the China
universities adopt this textbook for UGs atomic-physics teaching.
In average it is a 4-credits UG course for one academic semester
and most of the universities in China urge the lecturer of this
course to complete the teaching within 18 or 19  weeks with 4 or 5
(class) hours per week.
\newline To be specific, the textbook contents$^4$ are listed below :
\newline Chapter 1. Introduction to Atoms \newline Chapter 2.
Energy Levels and Radiation of Atoms \newline Chapter 3.
Fundamentals of Quantum Mechanics \newline Chapter 4. Alkali-metal
Atoms and Electron Spin
\newline Chapter 5. Multi-Electron Atoms \newline Chapter 6. Atoms
in Magnetic Field \newline Chapter 7. Shell Structure of  Atoms
\newline Chapter 8. X-ray \newline Chapter 9. Molecular Structure
and Molecular (Optical) Spectra \newline Chapter 10. Atomic Nuclei
\newline Chapter 11. Fundamental Particles \newline
Due to the limitation of time and the knowledge of UGs for each
semester, normally only around 8 chapters are covered during the
teaching. Note that most of the contents of this textbook (up to
Chapter 7) borrow those in Refs. 6-9 and relevant material from
Russia. There are homework assignments for UGs of this course and
the percentage is only 10\% for the total score (as UGs will copy
the solutions of the problems easily before submitting). The
primary verification of the understanding as well as learning of
atomic-physics for UGs from this course is through the final
examination or (written) test.\newline Two groups of UGs from
Xinjiang Normal University were tested on their knowledge acquired
by having attended a course in atomic physics. Fig. 1 gives the
exact distribution of questions on each chapter or main category
for a (written) final examination (for year-2 Han-people UGs).
There are 25 questions in total which are (i) single-choice
problem : 10 sets (each of the problem is graded by points 2);
(ii) filling-in problem : 7 sets (each of the problem is graded by
points ranging from  2 to 4); (iii) simplified-answer problem : 4
sets (each of the problem is graded by points 5); (iv) calculation
problem : 4 sets (each of the problem is graded by points 10).
There is limited percentage for Chapter 3 (related to quantum
mechanics$^{8-9}$) because all UGs have trouble in rapid
understanding the quantum-physics essentials before taking one
semester quantum-mechanics course from a short chapter contents of
the textbook. The latter situation is similar to other
countries$^{10-12}$. This can be easily found out during the class
teaching. For instance, students thought that the Heisenberg
uncertainty relation is too abstract. The other difficult topic is
related to the essence of spin!
\newline As the prerequisite background (say, physical
mathematics, electrodynamics, thermodynamics) is much more worse
for Other-people (mainly Uygur- and Kazak-people) UGs, the
detailed distribution of questions on each chapter or main
category for a (written) final examination (for year-3
Other-people UGs) is slightly  different  from that in Fig. 1.
This can be evidenced in Fig. 2. There are no (directly-linked)
questions on Chapter 3 : Fundamentals of Quantum Mechanics. Please
refer to Fig. 3 for similar details.
 Note that the maximal evaluation mark in China universities is
100/100. The difficulty characterization for both written
examinations is only of medium difficulty. The 60-points grading
means the passing an examination in China universities.
\section{Comparative Study between Han- and Other-people UGs}
We illustrate the distribution of (total) marks for the final
(written) examination of atomic-physics course taken by year-2
Han-people and year-3 Other-people UGs, respectively in Figs. 4
and 5. The comparison could be directly  figured out from these
two figures. The number of Han-people UGs passing the final
examination within the semester during which they attended the
lectures is 27 (8 failed) reaches 77\% (23\% failed). \newline
Meanwhile around 92\% (8\%) Other-people (mainly Uygur- and
Kazak-people) UGs passed (failed) the final examination within the
semester during which they attended the atomic-physics lectures.
The general trend was that the Mandarin (most of them are
Han-people) UGs considered their theoretical background above and
the practical comparable or below that of the Uygur and Kazak UGs.
A trend was that to most of the year-2 Han-people UGs the focus is
rather on understanding than on memorizing. However, as UGs still
didn't learn the course : Quantum Mechanics in details, their
understanding (i.e., the ability to derive some of the simple
formula in atomic physics) is limited. During the final
examination, those who have better memorizing skill normally can
get high scores! This is especially suitable to year-3
Other-people UGs since their Chinese as well as Mandarin are
already poor enough (thus they try to keep in mind everything for
the preparation of final (written) examination of atomic physics).
\section{Discussion}
We can further demonstrate some of the troubled details about the
learning of atomic physics before attending the quantum mechanics
course. For instance, consider this question : Try to obtain the
rotating frequency, linear velocity of electron in the first Bohr
orbit (orbiting a nucleus of hydrogen atom) (cf. Chapter 2 in Ref.
4). We can easily work out the solutions by noting the development
of atomic physics.\newline Firstly in 1911 Rutherford discovered
the nucleus. This then led to the idea of atoms consisting of
electrons in certain orbits in which the central forces are
provided by the Coulomb attraction to the positive nucleus. In
1913 Bohr produced his model for the atom. The key new elements of
the model are: \newline (a) The angular momentum $L$ of the
electron is quantized in units of $\hbar$ ($\hbar = h/2\pi$):
\newline \hspace*{5mm} $L = n \hbar$ ; where $n$ is an integer.
\newline (b) The atomic orbits are stable, and light is only
emitted or absorbed when the electron jumps \newline \hspace*{5mm}
from one orbit to another. \newline When Bohr made these
hypotheses in 1913, there was no justification for them other than
they were spectacularly successful in predicting the energy
spectrum of hydrogen. With the hindsight of quantum mechanics, we
now know why they work. The first assumption is equivalent to
stating that the orbit must correspond to a fixed number of de
Broglie wavelengths. For a circular orbit, this can be written:
\begin{equation}
  2\pi r =n \times \frac{h}{p}=n\times \frac{h}{mv},
\end{equation}
which can then be rearranged to give $L=m v r=n\times {h}/{2\pi}$.
With the final expression, we can directly
calculate the linear velocity ($v$) and then the rotating frequency
(set $r=a_1$, $a_1=5.29\times 10^{-11}$m :
the first Bohr orbit).\newline
The second assumption is a consequence of the fact that the Schr\"{o}dinger
equation
\begin{equation}
  (H_0 + V) \psi = E \psi
\end{equation}
leads to time-independent solutions (eigenstates) where the
simplest form of $H_0$ is
\begin{displaymath}
  H_0 \equiv  - \frac{\hbar^2}{2m}\frac{d^2}{dx^2},
\end{displaymath}
and $V\equiv V(x)$  is the possible potential. Furthermore, the
derivation of the quantized energy levels proceeds as follows.
Consider an electron orbiting a nucleus of mass $m_N$ and charge
$+Ze$. The central force is provided by the Coulomb force:
\begin{equation}
  F=\frac{m v^2}{r}=\frac{Z e^2}{4\pi \epsilon_0 r^2}.
\end{equation}
As with all two-body orbit  systems, the mass $m$ that enters the
formula is the reduced mass :
\begin{displaymath}
 \frac{1}{m}=\frac{1}{m_e}+\frac{1}{m_N}.
\end{displaymath}
The energy is given by the sum of kinetic energy and potential
energy
\begin{equation}
 E_n=\frac{m v^2}{2} -\frac{Z e^2}{4\pi \epsilon r}
    =\frac{m Z^2 e^4}{8 \epsilon^2 h^2 n^2}
\end{equation}
where we have used $v=(n h)/(m r 2\pi)$ and Eq. (3).  This can be
written as
\begin{displaymath}
  E_n =-\frac{R'}{n^2}
\end{displaymath}
with
\begin{equation}
  R'=\frac{m}{m_e}Z^2 R_{\infty},
\end{equation}
where $R_{\infty}$ is the Rydberg energy.
\newline All UGs
should memorize some of the results (say, quantized energy levels,
the Rydberg energy : $m_e e^4/(8\epsilon^2_0 h^2)=2.18\times
10^{-18}$J= 13.6 eV, the (effective) Rydberg constant) which can
be derived after learning the quantum mechanics. \newline Above
mentioned facts could explain why the year-3 Other-people UGs got
better scores than the year-2 Han-people after the final
examination of atomic physics. Other-people UGs prepared the test
as well as doing exercises mostly by memorizing. Note that here
the year-3 Other-people UGs were selected after their year-1
studies according to their better gradings of those courses they
have taken during the first year study at XJNU.
\newline The teaching of atomic physics is now plainly in a
unstable state. The interest of undergraduates for this discipline
is falling sharply, and the general level of achievement is poor,
as it is for specific abilities also (say, earlier knowledge of
quantum physics). The solutions that we, as lecturer or
instructors, have to grasp in order to overcome these
circumstances are$^{13}$: \newline (i) A flexible teaching system
for different-people UGs.
\newline (ii) The use while teaching of computer and communication
technologies: liable in themselves to render courses much more
attractive.
\newline (iii) Introduction of up-to-date themes, the study of
which would provide undergraduates with highly useful knowledge in
the field for which they are training. It is not normal for
courses in atomic physics taught in the late 2000s to stop the
learning at the level of knowledge that was the current in the
1940s.
%-------
 {\it
Acknowledgements.} The first author thanks the support of Xinjiang
Normal University.
%The second author thanks the support by the
%Starting Funds for 2009-Hebei-Normal-Univ. Scientific Researcher.

\newpage
%\subsubsection*{Acknowledgements}
%{\small  }
\psfig{file=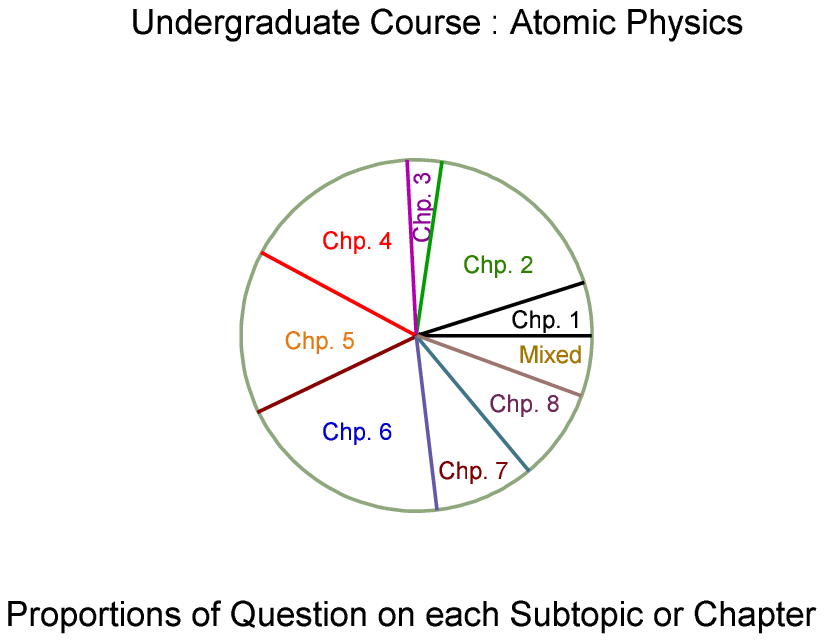,bbllx=0cm,bblly=12.0cm,bburx=18.5cm,bbury=26cm,rheight=9cm,rwidth=12cm,clip=}
%
%\vspace{2mm}
\begin{figure}[h]
\hspace*{6mm} Fig. 1 \hspace*{1mm} Distribution of problems for
the atomic-physics final examination for \newline \hspace*{7mm}
Han-people UGs.
\end{figure}

\newpage

\psfig{file=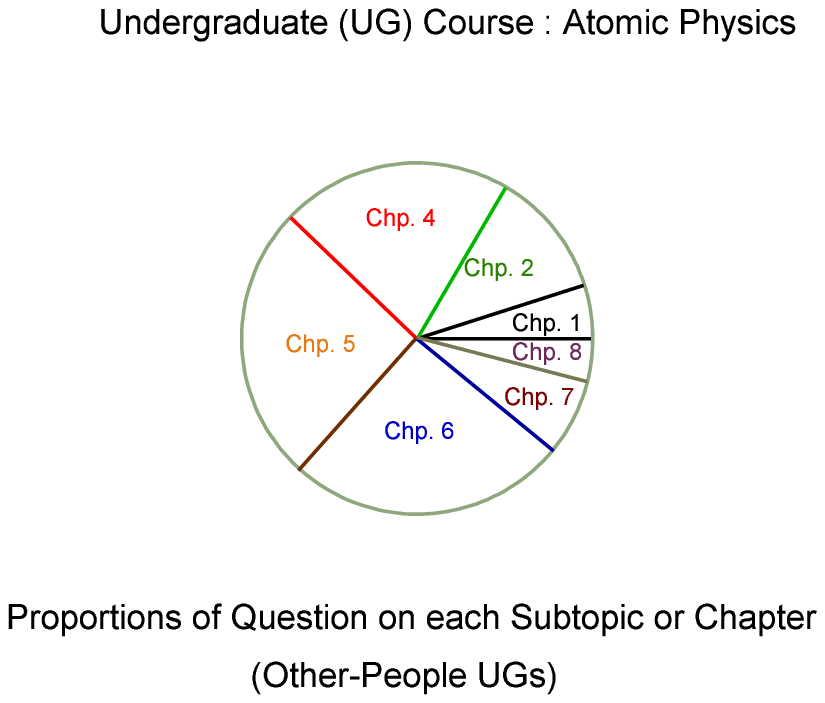,bbllx=0cm,bblly=12.0cm,bburx=18.5cm,bbury=26cm,rheight=9cm,rwidth=12cm,clip=}
%
%\vspace{2mm}
\begin{figure}[h]
\hspace*{6mm} Fig. 2 \hspace*{1mm} Distribution of problems for
the atomic-physics final examination for \newline \hspace*{7mm}
Other-people (like Uygur and Kazak) UGs.
\end{figure}

\newpage

\psfig{file=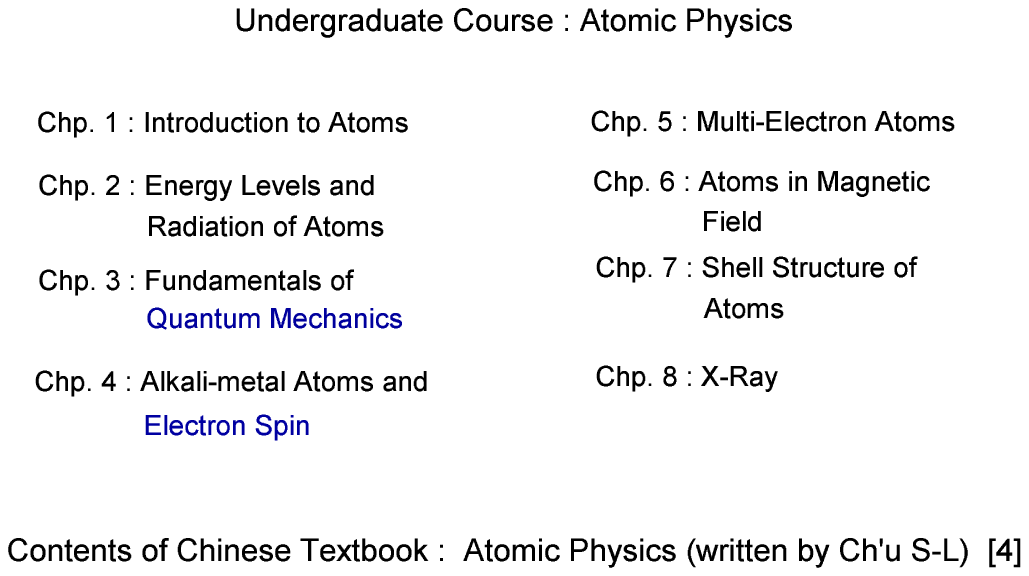,bbllx=-1cm,bblly=18.0cm,bburx=12cm,bbury=26.4cm,rheight=9cm,rwidth=12cm,clip=}
\begin{figure}[h]
\hspace*{6mm} Fig. 3 \hspace*{1mm} Outline of contents of the
textbook (in Chinese) for atomic physics$^4$. \newline
\hspace*{7mm} Dr. Ch'u S-L got PhD from Univ. Chicago around 1935
and was an expert in \newline \hspace*{7mm} high frequency spark
then$^5$.
\end{figure}

\newpage

\psfig{file=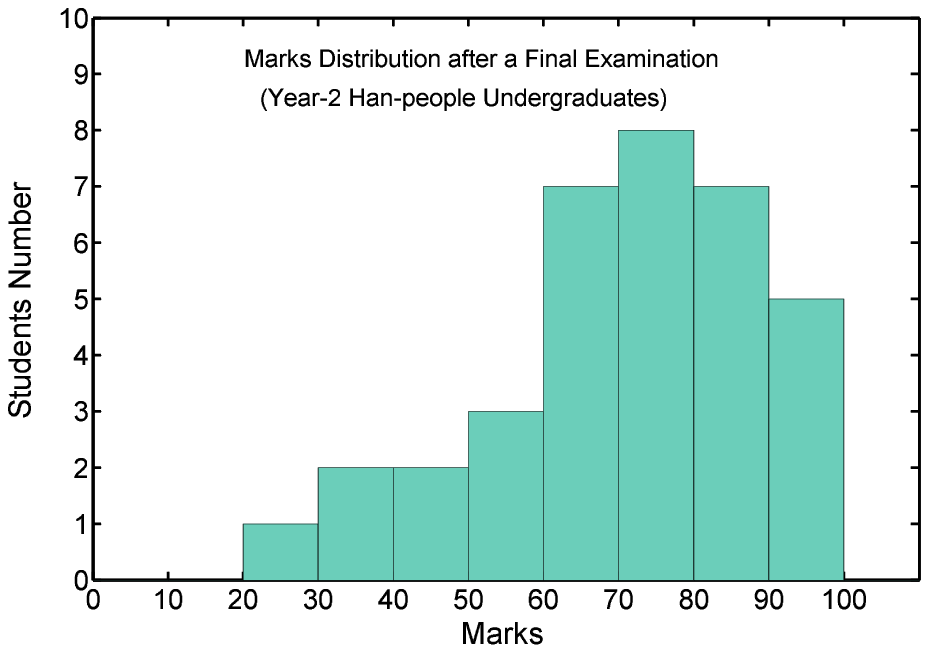,bbllx=0cm,bblly=18.0cm,bburx=18.5cm,bbury=26.4cm,rheight=9cm,rwidth=10cm,clip=}
%
%\vspace{2mm}
\begin{figure}[h]
\hspace*{6mm} Fig. 4 \hspace*{1mm} Distribution of marks for the
atomic-physics final examination for \newline \hspace*{7mm}
Han-people  UGs.
\end{figure}

\newpage

\psfig{file=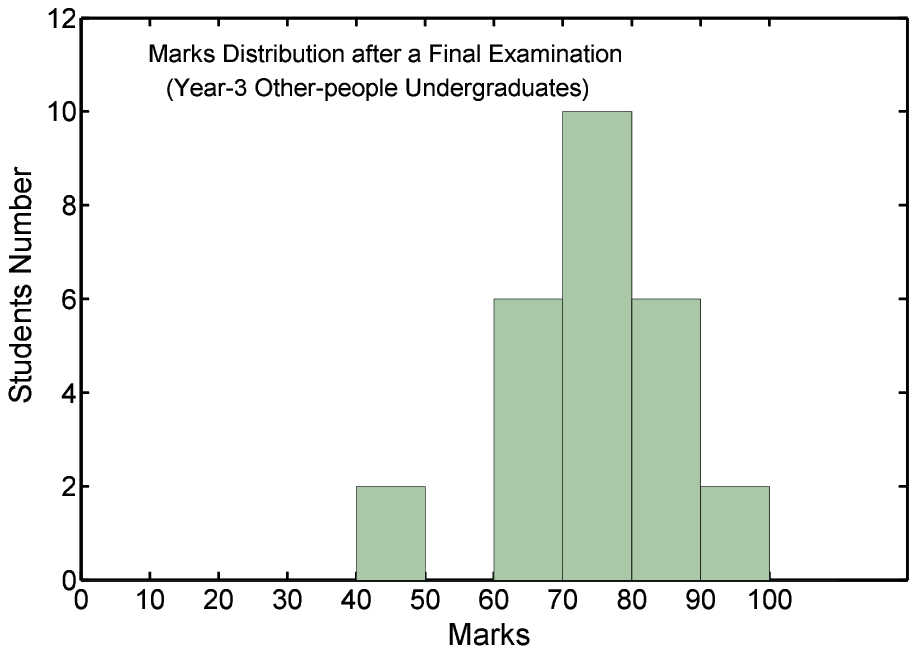,bbllx=0cm,bblly=18.0cm,bburx=18.5cm,bbury=26.4cm,rheight=9cm,rwidth=10cm,clip=}
%
%\vspace{2mm}
\begin{figure}[h]
\hspace*{6mm} Fig. 5 \hspace*{1mm} Distribution of marks for the
atomic-physics final examination for \newline \hspace*{7mm}
Other-people (like Uygur and Kazak) UGs.
\end{figure}

\end{document}